\documentclass[superscriptaddress,onecolumn,secnumarabic,
amssymb,amsmath,nobibnotes,aps,prd,showkeys,showpacs,nofootinbib]{revtex4}

\usepackage[latin1]{inputenc}
\usepackage{graphicx}
\usepackage[english]{babel}

\usepackage{amsmath}
\usepackage{amssymb}
\usepackage{amsfonts}
\usepackage{colordvi}
\usepackage{psfrag}
\usepackage{color}

\begin{document}
\def\cL{{\cal L}}
\def\be{\begin{equation}}
\def\ee{\end{equation}}
\def\bea{\begin{eqnarray}}
\def\eea{\end{eqnarray}}
\def\beq{\begin{eqnarray}}
\def\eeq{\end{eqnarray}}
\def\tr{{\rm tr}\, }
\def\nn{\nonumber \\}
\def\e{{\rm e}}

%
%


\def\bef{\begin{figure}}
\def\eef{\end{figure}}
\newcommand{\ans}{ansatz }
\newcommand{\eeqn}{\end{eqnarray}}
\newcommand{\bd}{\begin{displaymath}}
\newcommand{\ed}{\end{displaymath}}
\newcommand{\mat}[4]{\left(\begin{array}{cc}{#1}&{#2}\\{#3}&{#4}
\end{array}\right)}
\newcommand{\matr}[9]{\left(\begin{array}{ccc}{#1}&{#2}&{#3}\\
{#4}&{#5}&{#6}\\{#7}&{#8}&{#9}\end{array}\right)}
\newcommand{\matrr}[6]{\left(\begin{array}{cc}{#1}&{#2}\\
{#3}&{#4}\\{#5}&{#6}\end{array}\right)}
\newcommand{\cvb}[3]{#1^{#2}_{#3}}
\def\lsim{\raise0.3ex\hbox{$\;<$\kern-0.75em\raise-1.1ex
e\hbox{$\sim\;$}}}
\def\gsim{\raise0.3ex\hbox{$\;>$\kern-0.75em\raise-1.1ex
\hbox{$\sim\;$}}}
\def\abs#1{\left| #1\right|}
\def\simlt{\mathrel{\lower2.5pt\vbox{\lineskip=0pt\baselineskip=0pt
           \hbox{$<$}\hbox{$\sim$}}}}
\def\simgt{\mathrel{\lower2.5pt\vbox{\lineskip=0pt\baselineskip=0pt
           \hbox{$>$}\hbox{$\sim$}}}}
\def\unity{{\hbox{1\kern-.8mm l}}}
\newcommand{\eps}{\varepsilon}
\def\ep{\epsilon}
\def\ga{\gamma}
\def\Ga{\Gamma}
\def\om{\omega}
\def\omp{{\omega^\prime}}
\def\Om{\Omega}
\def\la{\lambda}
\def\La{\Lambda}
\def\al{\alpha}
\newcommand{\ov}{\overline}
\renewcommand{\to}{\rightarrow}
\renewcommand{\vec}[1]{\mathbf{#1}}
\newcommand{\vect}[1]{\mbox{\boldmath$#1$}}
\def\tm{{\widetilde{m}}}
\def\mcirc{{\stackrel{o}{m}}}
\newcommand{\Dm}{\Delta m}
\newcommand{\dm}{\varepsilon}
\newcommand{\tanb}{\tan\beta}
\newcommand{\nbar}{\tilde{n}}
\newcommand\PM[1]{\begin{pmatrix}#1\end{pmatrix}}
\newcommand{\up}{\uparrow}
\newcommand{\down}{\downarrow}
\def\omE{\omega_{\rm Ter}}
%

\newcommand{\Dsusy}{{susy \hspace{-9.4pt} \slash}\;}
\newcommand{\DCP}{{CP \hspace{-7.4pt} \slash}\;}
\newcommand{\mc}{\mathcal}
\newcommand{\gr}{\mathbf}
\renewcommand{\to}{\rightarrow}
\newcommand{\gtc}{\mathfrak}
\newcommand{\wh}{\widehat}
\newcommand{\br}{\langle}
\newcommand{\kt}{\rangle}


\def\lsim{\mathrel{\mathop  {\hbox{\lower0.5ex\hbox{$\sim$}
\kern-0.8em\lower-0.7ex\hbox{$<$}}}}}
\def\gsim{\mathrel{\mathop  {\hbox{\lower0.5ex\hbox{$\sim$}
\kern-0.8em\lower-0.7ex\hbox{$>$}}}}}

\def\nn{\\  \nonumber}
\def\de{\partial}
\def\brf{{\mathbf f}}
\def\bbf{\bar{\bf f}}
\def\bF{{\bf F}}
\def\bbF{\bar{\bf F}}
\def\bA{{\mathbf A}}
\def\bB{{\mathbf B}}
\def\bG{{\mathbf G}}
\def\bI{{\mathbf I}}
\def\bM{{\mathbf M}}
\def\bY{{\mathbf Y}}
\def\bX{{\mathbf X}}
\def\bS{{\mathbf S}}
\def\bb{{\mathbf b}}
\def\bh{{\mathbf h}}
\def\bg{{\mathbf g}}
\def\bla{{\mathbf \la}}
\def\bmu{\mathbf m }
\def\by{{\mathbf y}}
\def\bmu{\mbox{\boldmath $\mu$} }
\def\bsig{\mbox{\boldmath $\sigma$} }
\def\bunity{{\mathbf 1}}
\def\cA{{\cal A}}
\def\cB{{\cal B}}
\def\cC{{\cal C}}
\def\cD{{\cal D}}
\def\cF{{\cal F}}
\def\cG{{\cal G}}
\def\cH{{\cal H}}
\def\cI{{\cal I}}
\def\cL{{\cal L}}
\def\cN{{\cal N}}
\def\cM{{\cal M}}
\def\cO{{\cal O}}
\def\cR{{\cal R}}
\def\cS{{\cal S}}
\def\cT{{\cal T}}
\def\eV{{\rm eV}}

\title{External stability for Spherically Symmetric Solutions in Lorentz Breaking Massive Gravity}


\author{Andrea Addazi}

\affiliation{ Dipartimento di Fisica,
 Universit\`a di L'Aquila, 67010 Coppito AQ, Italy}
 
 \affiliation{Laboratori Nazionali del Gran Sasso (INFN), 67010 Assergi AQ, Italy}
 
 \author {Salvatore Capozziello}

\affiliation{Dipartimento di Fisica, Università
di Napoli {}``Federico II'', INFN Sez. di Napoli, Compl. Univ. di
Monte S. Angelo, Edificio G, Via Cinthia, I-80126, Napoli, Italy} 

\affiliation{INFN Sez. di Napoli, Compl. Univ. di
Monte S. Angelo, Edificio G, Via Cinthia, I-80126, Napoli, Italy}

\affiliation{Gran Sasso Science Institute 
(INFN), Viale F. Crispi 7, I-67100, L'Aquila, Italy.}

\date{\today}

\begin{abstract}

We discuss spherically symmetric solutions for point-like sources in Lorentz-breaking massive gravity theories. This analysis is valid for 
 St\"uckelberg's effective field theory formulation, for Lorentz Breaking Massive Bigravity and general extensions of gravity leading to an extra term $-Sr^{\gamma}$ added to the Newtonian potential. The approach consists in  analyzing  the stability of the geodesic equations, at the first order 
(deviation equation). 
The main result is a strong constrain in  the space of parameters 
of the theories. This  motivates  higher order  analysis of geodesic perturbations in order to understand if a class of spherically symmetric Lorentz-breaking massive gravity solutions, for  self-gravitating systems, exists.  Stable and phenomenologically acceptable solutions are discussed in the no-trivial case $S\neq 0$. 
\end{abstract}

\maketitle

\section{Introduction}

The issue of graviton mass is one of the most intriguing questions of theoretical physics. 
It is connected to other deep questions like the fact that  General Relativity is  an approximation of some fundamental  theory of gravity and 
Lorentz Invariance can be violated.

The question of graviton   mass  was firstly considered by  Fierz and Pauli (FP)
\cite{FP}. Despite of this intuition,  it was soon  clear  that this theory becomes unphysical: the modification of the Newtonian potentials is discontinuous in the limit of $m\rightarrow 0$ (very small graviton mass) with a large deviation of $25\%$ to the light deflection from the Sun \cite{Sun,Sun1,Sun2}. Alternatively, it was proposed that, in the full non-linear regime,  the discontinuity can be avoided \cite{Vai,8}. However the FP theory is also problematic at quantum level. In fact, the gauge symmetry is broken by the explicit mass $m$ term with cutoff $\sim (m^{4}M_{P})^{1/5}$. This scale is lower than the expected $(mM_{P})^{1/2}$ \cite{5}.
On the other hand, such shortcomings are avoided by considering a possible connection with the Lorentz-Invariance violation cited above. 
In fact it is interesting to consider Lorentz-breaking massive terms \cite{10}: this class of terms is free from ghosts at low and strong coupling scales. Mass terms, breaking the diffeomorphism invariance, have been considered by reintroducing the Goldstone field associated to the broken invariance (\cite{5,14} and for a review \cite{11}). Considering  Lorentz Breaking Massive Bigravity (LBMG) as an effective field theory,  the graviton mass is generated by
the interaction with a suitable set of St\"uckelberg fields. A set of four St\"uckelberg fields,  $\phi^{a}$ $(a=1,2,3,4)$, is introduced  by  transformations under  diffeomorphisms $\delta x^{\mu}=\zeta^{\mu}(x)$. Goldstone fields transform as scalars and they can realize a modified theory of gravity  at IR scales.  The result  is manifestly invariant under diffeomorphism. In particular, the Einstein-Hilbert action is extended with  potentials like $m^{2}\mathcal{V}(\mathcal{X},V^{i},S^{ij})$, with $\mathcal{X}=-g^{\mu\nu}\partial_{\mu}\phi^{0}\partial_{\nu}\phi^{0}$, $V^{i}=-g^{\mu\nu}\partial_{\mu}\phi^{i}\partial_{\nu}\phi^{0}$, $S^{ij}=-g^{\mu\nu}\partial_{\mu}\phi^{i}\partial_{\nu}\phi^{j}$, where $m$ is the graviton mass scale. Lorentz breaking is manifest in the action, but it can be considered as a  rotationally invariant potential $\mathcal{V}$. 
Bigravity could be another interesting approach to realize LBMG without ghosts and discontinuities \cite{12,exact, prado}. 
Such a theory is based on two coupled Einstein field equations derived on a Riemannian manifold with two conjugate metrics and two  geodesic structures \footnote{Other Lorentz breaking gravity models, generally known as Bumblebee gravity, are considered in 
\cite{K1}-\cite{B2014}. The strategy to break Lorentz symmetry is completely different in this case: the Einstein-Hilbert action is extended with 
an interaction term of the curvature tensor and a new vector field (see \cite{K1}). Then the vector field takes an expectation value
that spontaneously breaks Lorentz invariance. Another approach comes from DGP models. As pointed out in \cite{Dvali}, spontaneous Lorentz breaking could be induced by the expectation value of sources.}. Bigravity is also a theory with an intriguing phenomenology in galactic physics (see \cite{trygravity, curves} and \cite{Rossi}), strictly related to the so-called Mirror Theories \cite{Mirror1,Mirror2,Mirror3,Mirror4,Mirror5,Mirror6}. 

Assuming  spherical symmetry for the metric generated by a source, it is possible to write 
\be \label{ds}
ds^{2}=-dt^{2}J(r)+K(r)dr^{2}+r^{2}d\Omega^{2}
\ee
As it was shown in \cite{SB}, choosing a class of $\mathcal{V}$ as a particular polynomial combination of  $\omega_{n}=\rm Tr(W^{n})$,
one arrives to 
\be \label{Solution}
J(r)=1-\frac{2GM}{r}+\Lambda^{2}r^{2}-2GSr^{\gamma};\,\,\,K(r)=\frac{k_{0}}{J(r)}
\ee
where $M$ and $S$ are two integration constants, $G=1/16\pi M_{P}^{2}$ is the Newton constant.  $\Lambda^{2}$ is the "effective" cosmological constant; it is a particular combination of the parameters in the potential chosen in literature. The exponent $\gamma$ is another combination of the parameters. The details are discussed in \cite{SB} and we report the relevant aspects for our purposes in Appendix A. 

The last power law in (\ref{Solution}) is the new term coming from the gravity modification; $1/r$ and $r^{2}$, on the contrary, are also present in the Schwarzschild-de Sitter metric. 
The same solution of (\ref{Solution}) was previously found in LBMG for a certain class of interactions between the two metrics \cite{exact}. 

Recently the specific class of solutions discussed in \cite{SB}, with a solution like (\ref{Solution}), is discovered to be unstable if $S\neq 0$ \cite{Pilo2013}. A complete non-perturbative analysis shows the presence of a hidden ghost. On the other hand, this strongly depends on the specific choice of the interaction potential $\mathcal{V}$. 
The aim of this paper is more general. We will point out that   a class of ghost-free LBMG exists with a static-gravitational potential like (\ref{Solution}). 
For example {\it kinetic mixing terms}, between the metric and the Goldstone fields or the two metrics in bigravity, could enter in the action, eliminating ghosts. These terms are not considered in literature. This could enlarge the space of the parameters with other peculiar effects to explore. 
Alternatively, it could be possible to extend the Einstein-Hilbert kinetic term of LBMG as a local  $f(R)$ term (see \cite{fR} for a review 
of $f(R)$ theories). $f(R)$ gravity itself is an intriguing extension of General Relativity with a lot of phenomenological aspects in cosmology \cite{fRcosmology1, fRcosmology2, fRcosmology3, fRcosmology4, fRcosmology5, fRcosmology6, fRcosmology7}, galatic physics \cite{galacticfR1,galacticfR2, galacticfR3, galacticfR4, galacticfR5, galacticfR6,galacticfR7},
gravitational waves \cite{fRwaves1, fRwaves2} and neutron stars \cite{odintsov1,odintsov2}. In summary, it could be interesting to consider $f(R)$ extension with a Lorentz breaking potential. 

Then, it is also possible to consider non-local kinetic terms in LBMG. 
Recently  the possibility of a ghost-free massive bigravity with non-local terms
was proposed  \cite{Maggiore}. For implications on the static potential in non-local modification of gravity 
see also \cite{Conroy:2014eja}.
On the other hand, a more general and radical point of view about ghosts in Extend Theories of Gravity is considered in \cite{WavesGhost}:
it was proposed that ghosts in quantum gravity could not be dangerous if one considers a different interpretation of the quantum mechanics. This approach could be  tested by gravitational waves.

Here, we want to study the external stability of the trajectories in a LBMG metric (\ref{ds}) through {\it the geodesic stability condition}. The analysis would be "effective", without  specifying the particular model and its problems. The general question of ghosts is beyond the scope of this paper. 
For "external stability",  we want to study the geodesic stability out of the event horizon for black holes: obviously this condition is automatically satisfied for stars.
In other words, we are studying the geodesic structure for all classes of LBMG theories with a new terms like $\sim r^{\gamma}$ in addition to the 
Newtonian gravitational potential. One can  study this term for all the possible values of $S,\gamma$.
A first constrain comes from the convergence of the Komar integral \cite{Komar} (see  \cite{exact} and \cite{SB}).
In fact, the gravitational energy is 
\be \label{Komar}
\mathcal{E}=-\frac{1}{4\pi G}\int_{\partial S} d^{2}x \sqrt{h}v^{\mu}u^{\nu}\mathcal{W}_{\nu;\mu}
\ee 
with $S$ is the 3-surface, $\partial S_{\bar{t}}$ is boundary at fixed time $\bar{t}$; 
$v^{\mu}$ is the normal versor of S, $u^{\nu}$ is the normal versor of the boundary. 
The integration leads to the result \footnote{The integration can be performed into a boundary as a 2-sphere,
$\mathcal{W}=\partial/\partial t$ the Killing vector associated to the time direction, the integration cutoff is to the 
fixed radius $R$.}
\be \label{Komar2}
\mathcal{E}=m-S\gamma R^{\gamma+1}
\ee
In the limit $R\rightarrow \infty$, the integral converges to $m$ only for $\gamma<-1$.
This means that our analysis will be physically interesting for gravitational potential $\gamma<-1$. 
In the case $\gamma>-1$, solutions are  excluded  thanks to the classical Komar bound.

The basic idea for studying  the geodetic stability is the following: let us assume  to infinitesimally perturb a generic geodetic trajectory in the gravitational metric (\ref{ds}) as $x^{\mu}\rightarrow x^{\mu}+\delta x^{\mu}$. If the 4-deviation $\delta x^{\mu}(s)$ explodes exponentially as $\delta x(s)\sim e^{k s}$ ($k$ is a constant),
 we have to conclude that the trajectories around the solutions are unstable. A solution for a star (or a black hole), that cannot admit external stable circular (or quasi-circular) trajectories, is not phenomenologically acceptable. In Section 2, we discuss this analysis and the consequent constraints on the space of parameters for LBMG theories. In Section 3, we consider the implication 
 of the analysis for a particular class of models studied in literature. Conclusions are drawn in Section 4. 
 In Appendix A, we report technical details of the models considered in Section 3. 
 In our analysis we will assume that $\Lambda=0$ in asymptotically flat hypothesis. 

\section{The geodesic stability condition}
The trajectories, in the  gravitational field background, are described by the {\it geodesic equations}
\be \label{geo}
\frac{d^{2}x^{\lambda}}{ds^{2}}+\Gamma_{\mu\nu}^{\lambda}\frac{dx^{\mu}}{ds}\frac{dx^{\nu}}{ds}=0
\ee
with ${\displaystyle \frac{dx^{\mu}}{ds}}$, the 4-velocity, $s$  the affine parameter along the geodesic. If we perturb the geodesic as 
$x^{\mu}\rightarrow x^{\mu}+\delta x^{\mu}$, where $\delta x^{\rho}$ is the 4-deviation,
we obtain, as standard, the deviation equation 
\be \label{gd}
\frac{d^{2}\delta x^{\lambda}}{ds^{2}}+2\Gamma_{\mu\nu}^{\lambda}\frac{dx^{\mu}}{ds}\frac{d\delta x^{\lambda}}{ds}+\partial_{\rho}\Gamma_{\mu\nu}^{\lambda}\frac{dx^{\mu}}{ds}\frac{dx^{\nu}}{ds}\delta x^{\rho}=0\,.
\ee
If we insert this into (\ref{ds}), fixing the constant $k_{0}=1$ in (\ref{Solution}),  we have following geodesic equations:
\be \label{ge1}
\frac{d^{2}t}{ds^{2}}=0,\,\,\,\frac{1}{2}J(r)'\left(\frac{dt}{ds}\right)^{2}-r\left(\frac{d\phi}{ds}\right)^{2}=0,\,\,\,\frac{d^{2}\theta}{ds^{2}}=0,\,\,\,\frac{d^{2}\phi}{ds^{2}}=0\,.
\ee
The geodesic deviation, divided by components, is
\be \label{d1}
\frac{d^{2}\delta x^{0}}{ds^{2}}+\frac{J'(r)}{J(r)}\frac{dt}{ds}\frac{d\delta x^{1}}{ds}=0\,,
\ee
\begin{eqnarray}
\frac{d^{2}\delta x^{1}}{ds^{2}}+J(r)J'(r)\frac{dt}{ds}\frac{d\delta x^{0}}{ds}-2rJ(r)\frac{d\phi}{ds}\frac{d\delta x^{3}}{ds}+
 \label{d2}
\left[\frac{1}{2}\left(J'^{2}(r)+J(r)J''(r)\right)\left(\frac{dt}{ds}\right)^{2}-(J(r)+rJ'(r))\left(\frac{d\phi}{ds}\right)^{2}\right]\delta x^{1}=0\,,
\end{eqnarray}
\be \label{d3}
\frac{d^{2}\delta x^{2}}{ds^{2}}+\left(\frac{d\phi}{ds}\right)^{2}\delta x^{2}=0\,,
\ee
\be
\label{d3a}
\frac{d^{2}\delta x^{3}}{ds^{2}}+\frac{2}{r}\frac{d\phi}{ds}\frac{d\delta x^{1}}{ds}=0
\ee
where $J'(r)=dJ(r)/dr$.
We consider the circular orbit in the plane $\theta=\pi/2$, that is the $ds^{2}$,  defined as (\ref{ds}), gives 
\be \label{cond}
J(r)\left(\frac{dt}{ds}\right)^{2}-r^{2}\left(\frac{d\phi}{ds}\right)^{2}=1
\ee
and from this last equation and Eq.(\ref{ge1}),  we  obtain
\be \label{o1}
\left(\frac{d\phi}{ds}\right)^{2}=\frac{J'(r)}{r[2J(r)-rJ'(r)]},\,\,\,\left(\frac{dt}{ds}\right)^{2}=\frac{2}{2J(r)-rJ'(r)}
\ee

According to this result, it is possible to  eliminate the dependence on $s$ in the deviation equations, obtaining
\be \label{d11}
\frac{d^{2}\delta x^{0}}{d\phi^{2}}+\frac{J'(r)}{J(r)}\frac{dt}{d\phi}\frac{d\delta x^{1}}{d\phi}=0
\ee
\be \frac{d^{2}\delta x^{1}}{d\phi^{2}}+J(r)J'(r)\frac{dt}{d\phi}\frac{d\delta x^{0}}{d\phi}-2rJ(r)\frac{d\delta x^{3}}{d\phi}
 \label{d21}
+\left[\frac{1}{2}\left(J'^{2}(r)+J(r)J''(r)\right)\left(\frac{dt}{d\phi}\right)^{2}-(J(r)+rJ'(r))\right]\delta x^{1}=0\,,
\ee
\be \label{d31}
\frac{d^{2}\delta x^{2}}{ds^{2}}+\delta x^{2}=0\,,
\ee
\be
\label{d31a}
\frac{d^{2}\delta x^{3}}{ds^{2}}+\frac{2}{r}\frac{d\delta x^{1}}{ds}=0
\ee
The last two equations give    harmonic motions  meaning that   motion in the plane $\theta=\pi/2$ is stable.
Let us   now  insert the modified gravitational potential (\ref{Solution}) into (\ref{d11}) and  (\ref{d21}). It is
\be \label{exp1}
\frac{d^{2}\delta x^{0}}{d\phi^{2}}+\frac{2GM+2\gamma G S r^{\gamma+1}}{r-2GM+2GSr^{\gamma+1}}\frac{dt}{d\phi}\frac{d\delta x^{1}}{d\phi}=0
\ee

\be \label{exp2}
0=\frac{d^{2}\delta x^{1}}{d\phi^{2}}+\left(1-\frac{2GM}{r}+2GSr^{\gamma}\right) \left(\frac{2GM}{r^{2}}+2\gamma G S r^{\gamma-1}\right)
\frac{dt}{d\phi} \frac{d\delta x^{0}}{d\phi}-2(r-2GM+2GS r^{\gamma+1})\frac{d\delta x^{3}}{d\phi}
\ee
$$+\left\{ \frac{1}{2} \left[4G^{2} \left( \frac{M}{r^{2}}+\gamma S r^{\gamma-1}\right)^{2}+\left(1-\frac{2GM}{r}+2GS r^{\gamma} \right)\left(2G\gamma S(\gamma-1)r^{\gamma-2}-\frac{4GM}{r^{3}} \right) \right] \left(\frac{dt}{d\phi}\right)^{2}-[1+2(1+\gamma)GSr^{\gamma}] \right\}\delta x^{1}$$
One  can  insert  the harmonic solutions
\be \label{ansatz}
\delta x^{0}=\delta x^{0}_{0}e^{i\omega \phi},\,\,\,\delta x^{1}=\delta x^{1}_{0}e^{i\omega \phi},\,\,\,\delta x^{3}=\delta x^{3}_{0}e^{i\omega \phi}
\ee
where $\delta x^{0,1,3}_{0}$ are constants.  The conditions obtained are apparently non-trivial, 
but between these, the following relevant relation with the gravitational potential (\ref{Solution}) is
\be \label{co1}
r^{3+\gamma}+\frac{2S}{M}>0
\ee
assuming  $G=\hbar=c=1$.
Condition (\ref{co1}) is not so different from the standard  condition for the Reissner-Nordstr$\ddot{o}$m case.
Let us  assume $\gamma=-2$ and $\frac{2S}{M}=-\frac{q^{2}}{m}$, with $q,m$ the charge and the mass of the particle. 
This is exactly the case of a symmetric solution for a charged particle. 
Also in  the Reissner-Nordstr$\ddot{o}$m case, constraints can be reduced  to  \cite{RN1,RN2}:
\be \label{simpleEx1}
r-6m>0
\ee
and
\be \label{simpleEx2}
r-\frac{q^{2}}{m}>0
\ee
Condition (\ref{simpleEx2})  corresponds exactly  to (\ref{co1}) with $\gamma=-2$ and $\frac{2S}{M}=-\frac{q^{2}}{m}$:
$r+\frac{2S}{M}>0$.  In other words, the bound (\ref{co1}) passes this consistency check  with standard results of General Relativity. 
However, Eq. (\ref{co1}) excludes the large region of parameters  $S<0$ for $\gamma <-3$. 
In fact for $r\rightarrow \infty$, $r^{3+\gamma}|_{\gamma<-3}\rightarrow 0$ while $S$ remains constant:
$0>-\frac{2S}{M}=\rm const>0$ is clearly impossible. 
For $-3<\gamma< -1$, solutions are possible because now, for $r\rightarrow \infty$, $r^{3+\gamma}|_{\gamma>-3}\rightarrow \infty$,  this is higher than the constant $\frac{2S}{M}$. This is expected considering the case of  Reissner-Nordstr$\ddot{o}$m, that is stable under 
geodetic perturbations. 
In other words,  we have a bound from below that not all the solutions satisfy (and again this is true also  the Reissner-Nordstr$\ddot{o}$m case, with a bound depending on the charge and the mass).
On the other hand,  the class of solutions  $S>0$ are unconstrained from the condition \footnote{See also \cite{Arraut1,Arraut2} for a recent study on the geodesic stability in other contests.}  (\ref{co1}).

\section{Constraints on the Space of parameters for a class of LBMG}
As an example, we discuss the implications of the above analysis on a particular class of LBMG
(see Appendix A). 

In particular, in \cite{SB}, the space of the parameters, allowing the 
stability of a self-gravitating system (a star), was studied: they consider  bounds coming from the internal pressure and the gravitational mass.
In fact, considering the internal pressure in the star,  it is possible to put bounds to the graviton mass. The pressure must be positive for the stability. At the star center,  this condition corresponds to 
\be \label{prho}
\frac{p(0)}{\rho_{0}}\simeq \frac{GM_{0}}{2R}\left[1-\frac{16\mu^{2}R^{2}(11-2\gamma)}{5(2\gamma-1)(\gamma-4)(\gamma-2)}\right]>0\,.
\ee
On the other hand,  at the surface  of the star (i.e. at radius $R$),  the pressure is zero, so the derivative has to assume  negative values. 
Then, the gravitational mass must be positive:
\be \label{M}
M=M_{0}\left[1-\frac{8\mu^{2}R^{2}}{5(\gamma+1)(\gamma-2)}\right]>0
\ee
These two conditions, studied in \cite{SB}, constraint the space of the parameters. For $\gamma=-1$, the bound (\ref{M}) leads to a divergence, so this is excluded by the space of the parameters. However,  we are considering the case $\gamma<-1$. 

From the above geodesic stability condition, we have another bound
\be \label{Condition11}
S=\frac{24\mu^{2}M_{0}R^{1-\gamma}}{(\gamma-4)(\gamma+1)(2\gamma-1)(\gamma-2)}>0
\ee
 corresponding to $\mu^{2} R^{2}=\rm const>0$ with respect to $\gamma$. 
This means that  the space of parameters is strongly constrained, especially from the bounds $M>0$ and $S>0$, as shown in Fig.1.
\begin{figure}[t]
\centerline{\includegraphics [height=14cm,width=1\columnwidth]{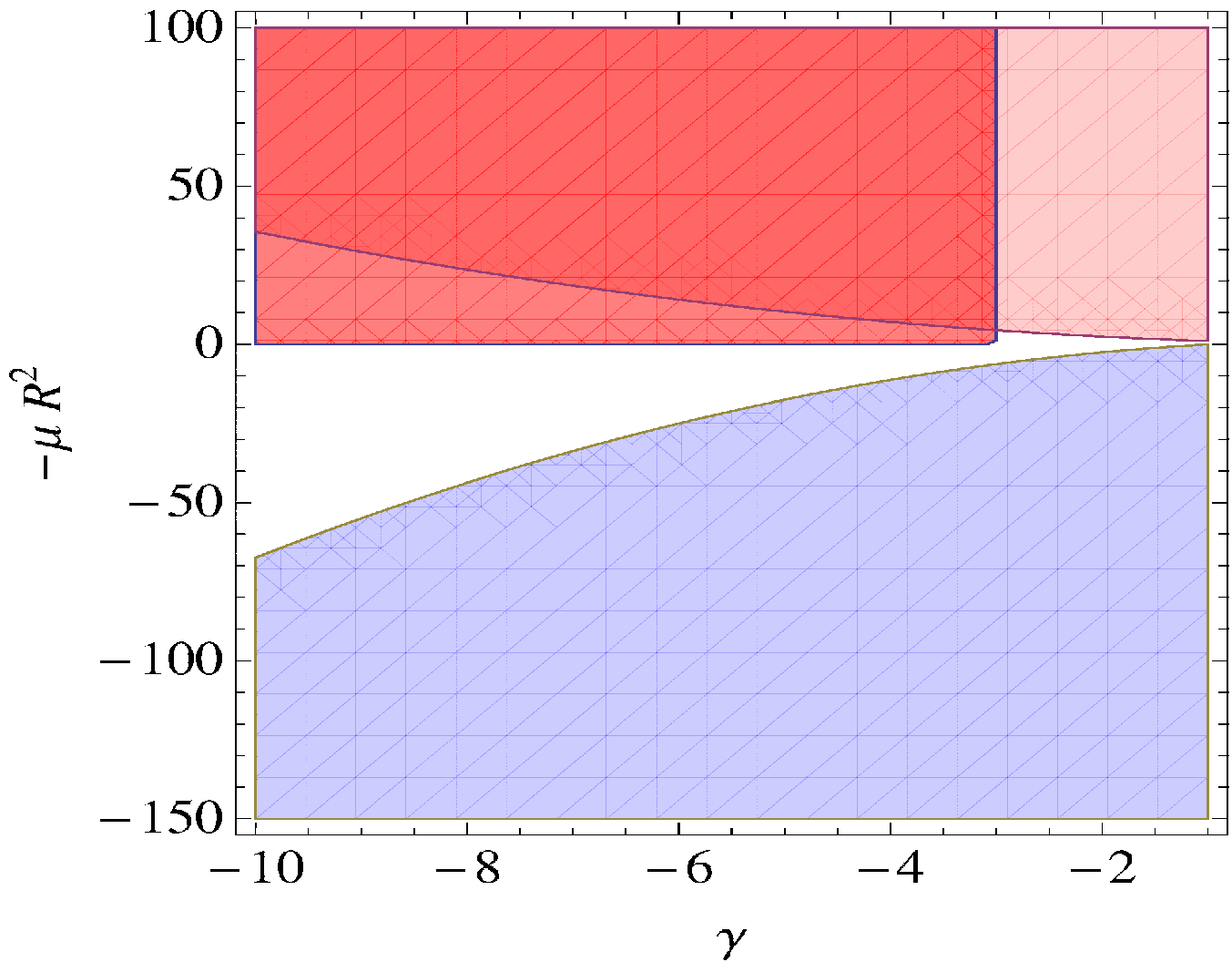} }
\vspace*{-1ex}
\caption{In the $(-\mu^{2}R^{2}, \gamma)$ plane, the white region is the subspace of allowed parameters. The regions in Pink and Blue are the Pressure and Mass bounds respectively, 
obtained in \cite{SB}. The new bound coming from the geodesic stability is reported in Red.
As a consequence, the space of the parameters is strongly constrained by these three bounds.}
\label{plot}   
\end{figure}
Assuming that all masses in the potential are of the
same order, limits on $\mu^{2}$ become limits on the graviton mass $m^{2}$: $m^{2}\sim |\mu|^{2}<\mathcal{O}(1)/R^{2}$. Assuming galaxies as the largest self-gravitating bound states implies $m<10^{-28\div 29}\, \rm eV$ for a typical galaxy size $R\sim 1\div 10\,\rm kpc$.

\section{Conclusions}
We analyzed the geodesic stability for spherically-symmetric solutions 
 in Lorentz-Breaking-Massive-Gravity, valid for St\"uckelberg, Bigravity and other LBMG classes leading to a new term $-Sr^{\gamma}$ in the static external potential. This analysis has an implication about the stability of self-gravitating systems (stars and black holes). 
 Clearly, not all the possible gauge interaction potentials were studied in the precedent literature, in the various classes of theories, essentially for a
 formal difficulty in the calculations. 
 This motivates a "bottom-up" approach starting from the gravitational static potential. 
 In fact, the gauge interaction potentials considered in literature, exclude very interesting 
 {\it kinetic mixing terms} between the Goldstone fields and the metric, or between the two metrics in bigravity.
 It is also practically unexplored the possibility to extend Einstein-Hilbert  kinetic terms of our metric to $f(R)$-like or other non-local theories of gravity. 
On the other hand, the interpretation of ghosts in quantum gravity could lead to ambiguities coming from the probabilistic formulation of quantum mechanics. 

 In our analysis, 
 we  assumed the case of asymptotically flat metrics, setting the effective cosmological term $\Lambda=0$.
 It was noted that $\gamma>-1$ leads to a divergent gravitational energy.
  As a consequence, our calculation assume the starting bound $\gamma<-1$.
 
The goal is a self-consistent physical restriction of  parameter space  for  static solutions.
 Calculations lead to exclude  metrics with $S<0$ for the non-standard term 
$\sim -Sr^{\gamma}$. In other words, the new term, proportional to $\sim r^{\gamma}$, {\it cannot be repulsive} if $\gamma<-3$. 
For $-3<\gamma<-1$ repulsive terms are possible, but with bounds from below depending on $S$ and $M$.
In Section 3, we have applied the bound from the geodetic analysis in a particular class of LBMG, explicitly showing how the bounds $S<0,\gamma<-3$ reduces the  parameter space. 

It is worth noticing  that our analysis is limited to the first order of  geodesic perturbation, a deeper analysis could constrain even
more the space of parameters for LBMG models. This strongly motivates a higher order geodesic analysis  to understand if a class of spherically symmetric LBMG solutions for stars and black holes exists in the non-trivial case $S\neq 0$.
This approach  could lead to important phenomenological implications.  It is worth stressing the fact that spherically symmetric solutions with charges or axial symmetry could have a larger parameter space, allowing  more "islands" of stable solutions. 

Finally, it is possible that spherically symmetric solutions for black holes develop inner blueshift  instabilities under electromagnetic or gravitational perturbations but this is not relevant for ordinary stars. Furthermore, strong gravity regimes could give rise to additional pressure terms capable of stabilizing very peculiar massive objects as discussed in\cite{odintsov1,odintsov2}.

\section{Appendix A: LBMG in St\"uckelberg's effective field theory formulation and Bigravity }

In this appendix, we sketch  the main aspects  of a class of LBMG 
realized as a St\"uckelberg's effective field theory and Bigravity. All these equations can be found in \cite{SB} and \cite{exact}. This section is necessary to understand the model dependent assumptions of the results showed in Section 3. 

Considering  LBMG as an effective field theory,  the graviton mass is generated by
the interaction with a suitable set of St\"uckelberg fields. A set of four St\"uckelberg fields.  $\Phi^{A}$ $(A=1,2,3,4)$, is introduced  by  transformations under  diffeomorphisms $\delta x^{\mu}=\zeta^{\mu}(x)$. Such fields transforms  as scalars and  can realize a  modified theory of gravity,  at IR scales,  which results  manifestly invariant under diffeomorphism.  Let us suppose to preserve the rotational invariance in the theory. 
The general action can be written as 
\be \label{S}
S=\int d^{4}x\sqrt{-g}M_{P}^{2}\{R+\mathcal{L}_{mat}+m^{2}\mathcal{V}(\mathcal{X},V^{i},S^{ij})\}
\ee
where $\mathcal{V}$ is a rotationally invariant potential and $\mathcal{X}=-g^{\mu\nu}\partial_{\mu}\Phi^{0}\partial_{\nu}\Phi^{0}$, $V^{i}=-g^{\mu\nu}\partial_{\mu}\Phi^{i}\partial_{\nu}\Phi^{0}$, $S^{ij}=-g^{\mu\nu}\partial_{\mu}\Phi^{i}\partial_{\nu}\Phi^{j}$. $m$ is the graviton mass scale. Lorentz breaking is manifest in the action. 
The Goldstone action can have additional symmetries
and can single out a particular phase for LBMG; we can consider $\mathcal{V}=\mathcal{V}(\mathcal{X},W^{ij})$ where $W^{ij}=S^{ij}-\mathcal{X}^{-1}V^{i}V^{j}$ \cite{14}. The Goldstone action is  invariant under $\Phi^{i}\rightarrow \Phi^{i}+\zeta^{i}(\Phi^{0})$.
The flat background is parametrizable as 
$\bar{g}_{\mu\nu}=\eta_{\mu\nu}$ and $\bar{\Phi}^{A}=(\alpha t,\beta x^{i})$, preserving rotation and breaking Lorentz boosts if $\alpha \neq \beta$. 

Assuming the spherical symmetry for the metric generated by a spherically symmetric source, it is possible to find out a set of coordinates in order to have  metric and  Goldstone fields in the simple form ($m_{1}=0$ phase). That is
\be \label{ds1}
ds^{2}=-dt^{2}J(r)+K(r)dr^{2}+r^{2}d\Omega^{2}
\ee
\be \label{Phi0}
\Phi^{0}=\alpha t+h(r),\,\,\,\Phi^{i}=\frac{x^{i}}{r}\phi(r)\,.
\ee
As it was shown in \cite{SB}, choosing the particular class of $\mathcal{V}$
\be \label{quite}
\mathcal{V}=[b_{0}+b_{1}\omega_{-1}+b_{2}(\omega_{-1}^{2}-\omega_{-2})+b_{3}(\omega_{-1}^{3}-3\omega_{-2}\omega_{-1}+2\omega_{-3})]\mathcal{X}^{-1}+
\ee
$$a_{0}+a_{1}\omega_{1}+a_{2}(\omega_{1}^{2}-\omega_{2})+a_{3}(\omega_{1}^{3}-3\omega_{2}\omega_{1}+2\omega_{3})\,,$$
with $\omega_{n}=\rm Tr(W^{n})$
one arrives to a Solution like (\ref{Solution}), with $\Lambda^{2}=\frac{1}{6}(12\bar{a}_{3}-6\bar{a}_{2}+\bar{a}_{0}-3\bar{b}_{1}+12\bar{b}_{2}-18\bar{b}_{3})$ is the "effective" cosmological constant; the exponent $\gamma$ is $\gamma =-2(2\bar{a}_{2}-6\bar{a}_{3}+\bar{b}_{1}-2\bar{b}_{2})/(\bar{a}_{1}-4\bar{a}_{2}+6\bar{a}_{3})$. 
The last power law in (\ref{Solution}) is the new term coming from the gravity modification, $1/r$ and $r^{2}$, on the contrary, are also present in the Schwarzschild-de Sitter metric. 
Considering the case $\Lambda=0$,  for $\gamma<-2$, the metric describes an asymptotically flat space. The masses of the fluctuations around the flat space are
$$m_{0}^{2}=\frac{3}{4}(\bar{a}_{1}-4\bar{a}_{2}+6\bar{a}_{3}-4\bar{b}_{2}-6\bar{b}_{3})m^{2},\,\,\,\,\,m_{1}^{2}=0$$
$$m_{2}^{2}=\frac{1}{2}(\bar{a}_{1}-2\bar{a}_{2}+\bar{b}_{1}-2\bar{b}_{2})m^{2}$$
$$m_{3}^{2}=\frac{1}{4}(\bar{a}_{1}-6\bar{a}_{3}-\bar{b}_{1}+8\bar{b}_{2}-18\bar{b}_{3})m^{2}$$
\be \label{mfluc}
m_{4}^{2}=\frac{1}{4}(\bar{a}_{1}-4\bar{a}_{2}+6\bar{a}_{3}-3\bar{b}_{1}+12\bar{b}_{2}-18\bar{b}_{3})m^{2}\,.
\ee
In this case, the mass of the graviton is $m_{2}$. 

The same solution of spherically symmetric solution is found in bigravity where Lorentz invariance is  spontaneously broken for a  large class of potentials \cite{exact}. 
In these models, we consider a theory with two interacting metrics: $g_{1}$ for  our observable sector and $g_{2}$ for another hidden sector. The two metrics are considered interacting through an effective potential. The resulting action is
\be \label{Lb}
S=\int d^{4}x[\sqrt{-g_{1}}(M_{pl1}^{2}+\mathcal{L}_{1})+\sqrt{-g_{2}}(M_{pl2}^{2}R_{2}+\mathcal{L}_{2})-4(g_{1}g_{2})^{1/4}\mathcal{V}(\mathcal{X})]\,.
\ee
 For symmetry, each rank-2 field is considered as coupled to its own matter field  with the respective Lagrangians
$\mathcal{L}_{1,2}$. For the sake of simplicity, we consider the interaction term without derivative couplings. Assuming  two metrics, the only possible combination is $\mathcal{X}_{\nu}^{\mu}=g_{1}^{\mu\alpha}g_{2\nu\alpha}$.  $\mathcal{V}$ is  a function of  the 4 independent scalar operators $\tau_{n}=\rm tr(\mathcal{X}^{n})$ $n=1,2,3,4$. We can include the cosmological term in our observable sector through a term in the effective potential as $\mathcal{V}_{\Lambda_{1}}=\Lambda_{1}q^{-1/4}$ with $q=\rm det \mathcal{X}=g_{2}/g_{1}$. The modified Einstein equations result
\be \label{E1}
M_{pl1}^{2}E_{1\nu}^{\mu}+Q_{1\nu}^{\mu}=\frac{1}{2}T_{1\nu}^{\mu}
\ee
\be \label{E2}
M_{pl2}^{2}E_{2\nu}^{\mu}+Q_{2\nu}^{\mu}=\frac{1}{2}T_{2\nu}^{\mu}
\ee
 where $Q_{1,2}$ are the effective energy-momentum tensors induced by the interaction
 \be \label{Q1}
 Q_{1\nu}^{\mu}=q^{1/4}[\mathcal{V}\delta_{\nu}^{\mu}-4(\mathcal{V}'\mathcal{X})_{\nu}^{\mu}]
 \ee
 \be \label{Q2}
 Q_{2\nu}^{\mu}=q^{-1/4}[\mathcal{V}\delta_{\nu}^{\mu}+4(\mathcal{V}'\mathcal{X})_{\nu}^{\mu}]
 \ee
with $(\mathcal{V}')_{\nu}^{\mu}=\partial \mathcal{V}/\partial \mathcal{X}_{\mu}^{\nu}$. 

The following Bianchi identities are satisfied:
\be \label{E12}
g_{1}^{\alpha \nu}\nabla_{1\alpha}E_{1\mu\nu}=\nabla_{1}^{\nu}E_{1\mu\nu}=0,\,\,\,\,\,g_{2}^{\alpha \nu}\nabla_{2\alpha}E_{2\mu\nu}=\nabla_{2}^{\nu}E_{2\mu\nu}=0\,.
\ee
They come from the invariance of the Einstein-Hilbert terms under  diffeomorphisms 
\be \label{deltag}
\delta g_{1\mu\nu}=2g_{1\alpha(\mu}\nabla_{\nu)}\epsilon^{\alpha},\,\,\,\,\,\delta g_{2\mu\nu}=2g_{2\alpha(\mu}\nabla_{\nu)}\epsilon^{\alpha}\,.
\ee
Being  the interaction term  also invariant, it is
$\nabla_{1,2}^{\nu}Q_{1,2\mu\nu}=0$ on shell for $g_{2,1}$. 

Regarding the asymptotic solutions, we expect that at infinity from the sources $g_{1,2}$ are maximally symmetric. 
Inserting 
\be \label{EoM}
M_{Pl 1,2}^{2}=E_{\mu\nu1,2}=\mathcal{K}_{1,2}g_{\mu\nu\,2}
\ee
with $-\mathcal{K}_{1,2}/4$, the constant scalar curvature of $g_{\mu\nu1,2}$, into the equations (\ref{E1}) and  (\ref{E2}),
we obtain 
\be \label{E11}
2\mathcal{V}+(q^{-1/4}\mathcal{K}_{1}+q^{1/4}\mathcal{K}_{2})=0
\ee
\be \label{E21}
8(\mathcal{V}'\mathcal{X})_{\nu}^{\mu}+\delta_{\nu}^{\mu}(q^{1/4}\mathcal{K}_{2}-q^{-1/4}\mathcal{K}_{1})=0\,.
\ee
Assuming asymptotically flat spaces $\mathcal{K}_{1,2}=0$, (\ref{E11}) and (\ref{E21}) are reduced to $\mathcal{V'}_{\mu}^{\nu}=0$ and $\mathcal{V}=0$. So, assuming that rotational symmetry is preserved and   the same signature for the two metrics (without  any 'twist' of coordinates), the bi-flat background can be written in the form
\be \label{g1}
\bar{g}_{1\mu\nu}=\eta_{\mu\nu}=\rm diag(-1,1,1,1)\,,
\ee
\be \label{g2}
\bar{g}_{2\mu\nu}=\omega^{2}\rm diag(-c^{2},1,1,1)\,,
\ee
where $\omega$ is the relative conformal factor, $c$ parametrizes the speed of light in sector $2$.
If $c=1$, the two metrics are linearly dependent and we can  locally diagonalize both of them by a  coordinate transformation. This is the case of Lorentz Invariant Bigravity. On the other hand, in the case  $c\neq 1$,  this  simultaneous diagonalization is not possible and then  the Lorentz symmetry is spontaneously broken. 

Considering the family of potentials 
\be \label{familyV}
\mathcal{V}=\alpha_{0}\mathcal{V}_{1}+\alpha_{1}\mathcal{V}_{2}+\alpha_{3}\mathcal{V}_{3}+\beta_{1}\mathcal{V}_{-1}+\beta_{2}\mathcal{V}_{-2}+\beta_{3}\mathcal{V}_{-3}+\beta_{4}\mathcal{V}_{-4}+q^{-1/4}\Lambda_{1}+q^{1/4}\Lambda_{2}
\ee
where we introduced the combinations of the scalars $\tau_{n}=\rm tr(\mathcal{X}^{n})$
\be \label{v0}
\mathcal{V}_{0}=\frac{1}{24|g_{2}|}(\epsilon\epsilon g_{2}g_{2}g_{2}g_{2})=1=\frac{1}{24q}(\tau_{1}^{4}-6\tau_{2}\tau_{1}^{2}+8\tau_{1}\tau_{3}+3\tau_{2}^{2}-6\tau_{4})
\ee
\be \label{v1}
\mathcal{V}_{1}=\frac{1}{6|g_{2}|}(\epsilon\epsilon g_{2}g_{2}g_{2}g_{1})=\tau_{-1}=\frac{1}{6q}(\tau_{1}^{3}-3\tau_{2}\tau_{1}+2\tau_{3})
\ee
\be \label{v2}
\mathcal{V}_{2}=\frac{1}{2|g_{2}|}(\epsilon\epsilon g_{2}g_{2}g_{1}g_{1})=(\tau_{-1}^{2}-\tau_{-2})=q^{-1}(\tau_{1}^{2}-\tau_{2})
\ee
\be \label{v3}
\mathcal{V}_{3}=\frac{1}{|g_{2}|}(\epsilon\epsilon g_{2}g_{1}g_{1}g_{1})=(\tau_{-1}^{3}-3\tau_{-2}\tau_{-1}+2\tau_{-3})=6q^{-1}\tau_{1}
\ee
\be \label{v4}
\mathcal{V}_{4}=\frac{1}{|g_{2}|}(\epsilon\epsilon g_{1}g_{1}g_{1}g_{1})=(\tau_{-1}^{4}-6\tau_{-2}\tau_{-1}^{2}+8\tau_{-1}\tau_{-3}+3\tau_{-2}^{2}-6\tau_{-4})=24q^{-1}
\ee
where $\mathcal{V}_{-n}=\mathcal{V}(\mathcal{X}\rightarrow \mathcal{X}^{-1})$;
we have a solution as (\ref{Solution}) plus  the second metric:
\be \label{dsg2}
ds_{2}^{2}=-Cdt^{2}+Adr^{2}+2Ddtdr+Bd\Omega^{2}
\ee
with
\be \label{coe2}
C=c^{2}\omega^{2}\left[1-\frac{2Gm_{2}}{\kappa r}+\mathcal{K}_{2}r^{2}\right]-\frac{2G}{c\omega^{2}\kappa}Sr^{\gamma},\,\,\,\,\,D^{2}+AC=c^{2}\omega^{4}
\ee
\be \label{coe3}
B=\omega^{2}r^{2},\,\,\,\,A=\omega^{2}\frac{\tilde{J}-\tilde{C}-\tilde{J}\tilde{S}r^{\gamma-2}}{\tilde{J}^{2}}
\ee

The forms of  $M,S$ in (\ref{Solution}) can be obtained by  matching the exterior solution with an  interior star  solution as  obtained in \cite{SB}.  Assuming the mass density of the object as a constant, the following  conditions
\be \label{Matching}
M=M_{0}\left[1-\frac{8\mu^{2}R^{2}}{5(\gamma+1)(\gamma-2)}\right],\,\,\,S=\frac{24\mu^{2}M_{0}R^{1-\gamma}}{(\gamma-4)(\gamma+1)(2\gamma-1)(\gamma-2)}
\ee
\be \label{mu2}
\mu^{2}=m_{2}^{2}\frac{3m_{4}^{4}-m_{0}^{2}(m_{2}^{2}-4m_{3}^{2})}{m_{4}^{4}-m_{0}^{2}(m_{2}^{2}-m_{3}^{2})},\,\,\,M_{0}=\frac{4}{3}\pi R^{3}\rho_{0}
\ee
hold.
As a consequence,  the bare mass $M_{0}$ is renormalized by the presence of the  extra-term  in $ S$ coming from modified gravity. 
Clearly, $M$ is the mass and $S$ depends on the size of the self-gravitating body.
The mass shift $\Delta M=M-M_{0}$ and $S$ will have the sign of $\mu^{2}$ in the case $\gamma<-1$. This means that  the corrections are proportional to the physical size of the body. For $\mu^{2}>0$, we have positive corrections $M>M_{0}$;
for $\mu^{2}<0$ we have negative  corrections. The parameter $\mu^{2}$ is proportional to the graviton mass scale $m^{2}$; the same holds for $S$ and $\Delta M$; in the limit of $m\rightarrow 0$,  the deviations go to zero and LBMG theory converges to General Relativity. 

Note that as a consequence of our analysis in Section 2, positive fluctuations are the only one permitted:
the bound $S>0$ excludes negative fluctuations.

\vspace{1cm} 

{\large \bf Acknowledgments} 
\vspace{3mm}

We would like to thank Denis Comelli, Luigi Pilo and Fabrizio Nesti for interesting discussions and important remarks on these subjects. SC is supported by INFN ({\it iniziative specifiche} TEONGRAV and QGSKY).

\end{document}